\documentclass{PoS} 

\usepackage{amsmath,amsthm}

\title{The spectral function in a strongly coupled, thermalising CFT} 

\ShortTitle{The spectral function in a strongly coupled, thermalising CFT} 

\author{Alice Bernamonti\\ 
Instituut voor Theoretische Fysica, KU Leuven,\\
Celestijnenlaan 200D, B-3001 Leuven, Belgium\\
E-mail: \email{alice@fys.kuleuven.be}} 

\author{Ben Craps\\ 
Theoretische Natuurkunde, Vrije Universiteit Brussel, and International Solvay Institutes,\\
Pleinlaan 2, B-1050 Brussels, Belgium\\
E-mail: \email{ben.craps@vub.ac.be}} 

\author{\speaker{Joris Vanhoof}\\
Theoretische Natuurkunde, Vrije Universiteit Brussel, and International Solvay Institutes,\\
Pleinlaan 2, B-1050 Brussels, Belgium\\
E-mail: \email{joris.vanhoof@vub.ac.be}} 

\abstract{In relation to the fluctuation-dissipation theorem, we discuss a time-dependent notion of spectral function and effective temperature. Extending recent results from \cite{Balasubramanian:2012tu}, we work out these quantities in a two-dimensional thermalising CFT dual to AdS$_3$-Vaidya spacetime which interpolates between a black brane geometry at early times and a higher temperature black brane at late times. The computation is carried out in the gravitational holographic dual in the geodesic approximation and using a non-standard analytic continuation. } 

\FullConference{ Proceedings of the Corfu Summer Institute 2012 \\ 
September 8-27, 2012\\ 
Corfu, Greece} 

\begin{document} 

%%%%%%%%%%%%%%%%%%%%%%%%%%%%%%%%%%%%%%%%%%%%%%%%%%%%%%%%%%%%%%%%%%%%%%%%%

\section{Introduction and summary}

Strongly coupled far-from-equilibrium systems appear in many areas of physics and their theoretical study is notoriously difficult. The gauge/gravity duality is a novel tool to study certain classes of strongly coupled field theories, and its extension to far-from-equilibrium situations is a very active area of current research. In order to extract useful field theory information from computations in the dual gravity theory, it is important to identify field theory observables that are meaningful in far-from-equilibrium states, as well as prescriptions for how to compute them in the gravity dual.

In the recent paper \cite{Balasubramanian:2012tu}, to which we refer for a more elaborate introduction as well as references to the literature, our collaborators and we studied a notion of time-dependent spectral functions and occupation numbers (based on Wigner transforms of two-point functions) in a simple holographic model (see  \cite{Chesler:2011ds,Banerjee:2012uq} for related work in other holographic models). Starting with the vacuum state of a strongly coupled 2d conformal field theory, the sudden, homogeneous injection of energy and its subsequent thermalization was modeled by an AdS$_3$-Vaidya geometry that interpolates between pure AdS at early times and a black brane geometry at late times. In order to compute two-point functions of high-dimension operators, we used a geodesic approximation \cite{Balasubramanian:1999zv}. For timelike separations of the operators, it was necessary to use either a non-standard Euclidean continuation of the bulk spacetime, or to use complex geodesics, and both procedures gave the same result. We also worked out an approach that went beyond the geodesic approximation, but will not discuss it here.

In the present contribution, we extend some of the results of \cite{Balasubramanian:2012tu} in two ways. First, instead of starting with the vacuum, we start with an initial thermal state. After a sudden, homogeneous injection of energy, the system will evolve towards a thermal state with higher temperature. Second, as in \cite{Chesler:2011ds}, we extract from our results a time-dependent effective temperature, which evolves in a non-monotonic way from the initial to the final temperature.

In section~\ref{Time-dependent two-point functions}, we review the notions of time-dependent spectral function \cite{Banerjee:2012uq, Balasubramanian:2012tu} and time-dependent effective temperature \cite{Chesler:2011ds}. In section~\ref{Setup of the model}, we present the AdS-Vaidya model that describes the evolution from a black brane with an initial temperature to a black brane with a larger final temperature via the injection of a shell of null dust. In this setting, section~\ref{Geodesic approximation of two-point functions} describes the computation of retarded and time-ordered two-point functions using a geodesic approximation and a non-standard continuation to Euclidean signature. In section~\ref{Spectral function and effective temperature in thermalising CFT}, these two-point functions are used to extract time-dependent spectral functions and effective temperatures, which are shown to interpolate between the initial and final equilibrium values.

%%%%%%%%%%%%%%%%%%%%%%%%%%%%%%%%%%%%%%%%%%%%%%%%%%%%%%%%%%%%%%%%%%%%%%%%%

\section{A notion of time-dependent spectral function and effective temperature}\label{Time-dependent two-point functions}

%%%%%%%%%%%%%%%%%%%%%%%%%%%%%%%%%%%%%%%%%%%%%%%%%%%%%%%%%%%%%%%%%%%%%%%%%

We probe the thermalisation of the field theory with time-ordered (Feynman) two-point functions,
\begin{eqnarray}
iG_{F}(t_{2},x_{2};t_{1},x_{1})&\equiv&\langle T\{\mathcal{O}(t_{2},x_{2})\mathcal{O}(t_{1},x_{1})\}\rangle \nonumber \\
&\equiv&\theta(t_{2}-t_{1})\langle\mathcal{O}(t_{2},x_{2})\mathcal{O}(t_{1},x_{1})\rangle+\theta(t_{1}-t_{2})\langle\mathcal{O}(t_{1},x_{1})\mathcal{O}(t_{2},x_{2})\rangle\,,
\end{eqnarray}
and retarded two-point functions (which are real in position space)
\begin{equation}
iG_{R}(t_{2},x_{2};t_{1},x_{1})\equiv\theta(t_{2}-t_{1})\langle[\mathcal{O}(t_{2},x_{2}),\mathcal{O}(t_{1},x_{1})]\rangle\,.
\end{equation}
From these definitions, using the hermiticity of the operator $\mathcal{O}(t,x)$, we can deduce the relation
\begin{equation}\label{eq:RelationFeynmanRetarded}
G_{R}(t_{2},x_{2};t_{1},x_{1})=\theta(t_{2}-t_{1})\left[G_{F}(t_{2},x_{2};t_{1},x_{1})+\left(G_{F}(t_{2},x_{2};t_{1},x_{1})\right)^{*}\right]\,,
\end{equation}
which we will employ later on. When the field theory is in equilibrium, the system is translation invariant in space and time. The correlators are then functions of the differences only, {\it i.e.}
\begin{equation}
G_{R,F}(t_{2},x_{2};t_{1},x_{1})=G_{R,F}(t_{2}-t_{1},x_{2}-x_{1};0,0)\equiv G_{R,F}(t,x)\,,
\end{equation}
with $t=t_{2}-t_{1}$ and $x=x_{2}-x_{1}$,  and their Fourier transforms are given by
\begin{equation}
G_{R,F}(\omega,k)=\int_{-\infty}^{+\infty}\mbox{d}t\int_{-\infty}^{+\infty}\mbox{d}x\,e^{i\omega t}e^{-ikx}G_{R,F}(t,x)\,.
\end{equation}
In momentum space, the retarded correlation function is in general complex valued, and its imaginary part defines the spectral function,
\begin{equation}\label{eq:DefSpectralFunction}
\rho(\omega,k)=-2\,\mbox{Im}\,G_{R}(\omega,k)\,.
\end{equation}
When the system is at thermal equilibrium, the retarded and time-ordered correlation functions are related by the Fluctuation-Dissipation Theorem,
\begin{equation}
\left[1+2n(\omega)\right]\rho(\omega,k)=-2\,\mbox{Im}\,G_{F}(\omega,k)\,,
\end{equation}
where the occupation number equals the Bose-Einstein distribution: $n(\omega)=(e^{\omega/\theta}-1)^{-1}$ for some temperature $\theta$. Note that using the previous expressions, we have
\begin{equation}
\theta=\frac{1}{2}\left.\mbox{Im}\,G_{F}(\omega)\left(\frac{\partial\mbox{Im}\,G_{R}(\omega)}{\partial\omega}\right)^{-1}\right|_{\omega=0}\,,
\end{equation}
where
\begin{equation}
G_{R,F}(\omega)=\int_{-\infty}^{+\infty}\frac{\mbox{d}k}{2\pi}\,G_{R,F}(\omega,k)=\int_{-\infty}^{+\infty}\mbox{d}t\,e^{i\omega t}G_{R,F}(t,x=0)
\end{equation}
is the momentum average of the correlators.

%%%%%%%%%%%%%%%%%%%%%%%%%%%%%%%%%%%%%%%%%%%%%%%%%%%%%%%%%%%%%%%%%%%%%%%%%

In a thermalising field theory, time translational invariance is generically broken. Therefore as in \cite{Balasubramanian:2012tu} we introduce an average time $T$ and a relative time $t$ by
\begin{equation}
\left\{\begin{array}{ll}
T=\frac{t_{1}+t_{2}}{2} \\
t=t_{2}-t_{1}
\end{array}\right.
\qquad\Leftrightarrow\qquad
\left\{\begin{array}{ll}
t_{1}=T-\frac{t}{2} \\
t_{2}=T+\frac{t}{2}
\end{array}\right.\,,
\end{equation}
and we define time-dependent correlators by keeping $T$ fixed and Fourier transforming with respect to the relative time $t$
\begin{equation}
G_{R,F}(\omega,T,k)=\int_{-\infty}^{+\infty}\mbox{d}t\int_{-\infty}^{+\infty}\mbox{d}x\,e^{i\omega t}e^{-ikx}G_{R,F}(t,T,x)\,,
\end{equation}
and
\begin{equation}
G_{R,F}(\omega,T)=\int_{-\infty}^{+\infty}\frac{\mbox{d}k}{2\pi}\,G_{R,F}(\omega,T,k)=\int_{-\infty}^{+\infty}\mbox{d}t\,e^{i\omega t}G_{R,F}(t,T,x=0)\,.
\end{equation}
We can then define a time-dependent spectral function
\begin{equation}\label{eq:DefTimeDepSpectralFunction}
\rho(\omega,T,k)=-2\,\mbox{Im}\,G_{R}(\omega,T,k)\,,
\end{equation}
and a time-dependent temperature
\begin{equation}\label{eq:DefTimeDepTemperature}
\theta(T)=\frac{1}{2}\left.\mbox{Im}\,G_{F}(\omega,T)\left(\frac{\partial\mbox{Im}\,G_{R}(\omega,T)}{\partial\omega}\right)^{-1}\right|_{\omega=0}\,.
\end{equation}
This notion of effective temperature was also considered in \cite{Chesler:2011ds} to assess thermalisation following an anisotropic deformation of the boundary.

%%%%%%%%%%%%%%%%%%%%%%%%%%%%%%%%%%%%%%%%%%%%%%%%%%%%%%%%%%%%%%%%%%%%%%%%%

\section{Setup of the model}\label{Setup of the model}

\begin{figure}[!h]
\centering
\includegraphics[width=8cm]{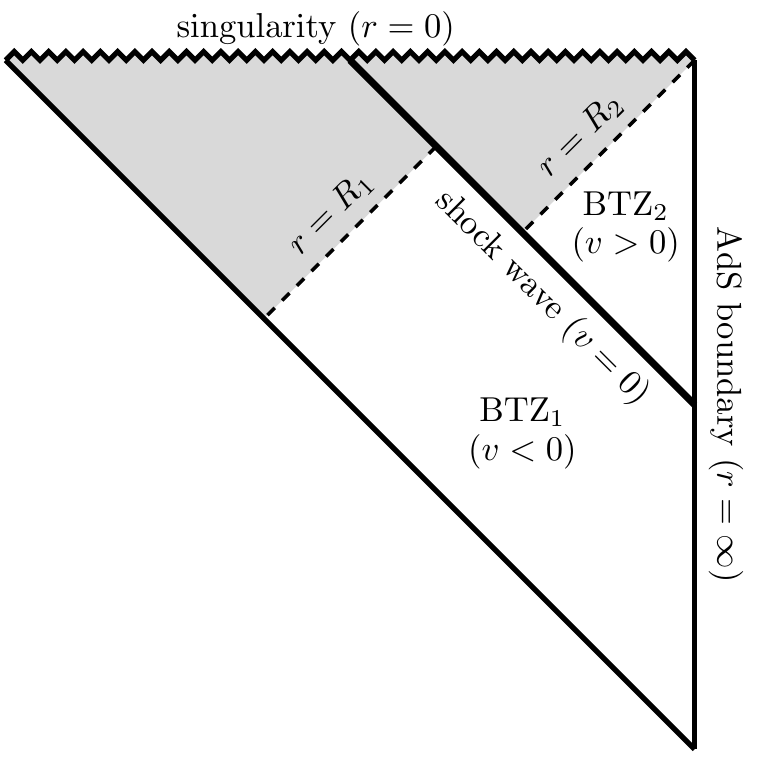}
\caption{\emph{The causal structure of the thin shell AdS-Vaidya spacetime.
}}\label{fig:penrose}
\end{figure}

As a holographic model for a thermalising field theory, we consider a three-dimensional, thin shell AdS-Vaidya spacetime. For $R_{2}>R_{1}>0$, it has the metric
\begin{equation}\label{eq:Vaidyametric}
ds^{2}=-[(r^{2}-R_{1}^{2})-\theta(v)(R_{2}^{2}-R_{1}^{2})]dv^{2}+2dvdr+r^{2}dx\,.
\end{equation}
The spacetime structure is depicted in Fig.~\ref{fig:penrose} and it represents an infalling shock wave of `null' dust in a black brane background that collapses to form a heavier black brane. The coordinate $v$ is an Eddington-Finkelstein coordinate such that $v=0$ corresponds to the location of the shock wave. The coordinate $r$ is a radial coordinate such that $r=\infty$ is the planar boundary of the asymptotically AdS spacetime and $r=0$ is the black brane singularity.

The metric (\ref{eq:Vaidyametric}) solves Einstein's equations with a negative cosmological constant and a delta source stress-energy tensor on the shock wave. Outside the shock wave ($v>0$), the metric is given by the BTZ solution with radius $R_{2}$
\begin{equation}\label{eq:BTZ2metric}
ds^{2}=-(r^{2}-R_{2}^{2})dt^{2}+\frac{dr^{2}}{r^{2}-R_{2}^{2}}+r^{2}dx
\qquad\mbox{with}\qquad
t=v-\frac{1}{2R_{2}}\ln\left|\frac{r-R_{2}}{r+R_{2}}\right|\,,
\end{equation}
while inside ($v<0$), we recover the BTZ metric with radius $R_{1}$
\begin{equation}\label{eq:BTZ1metric}
ds^{2}=-(r^{2}-R_{1}^{2})dt^{2}+\frac{dr^{2}}{r^{2}-R_{1}^{2}}+r^{2}dx
\qquad\mbox{with}\qquad
t=v-\frac{1}{2R_{1}}\ln\left|\frac{r-R_{1}}{r+R_{1}}\right|\,.
\end{equation}
The dual picture of this spacetime on the boundary CFT is that of a field theory that is initially in equilibrium at a temperature $\theta_{1}=\frac{R_{1}}{2\pi}$, which corresponds to the Hawking temperature of the initial black brane. After a sudden injection of energy at the time $t=0$, there is a period of thermalisation until the field theory reaches a new equilibrium at a higher temperature $\theta_{2}=\frac{R_{2}}{2\pi}$. 

We will now probe the equilibration dynamics of this holographic model using a geodesic approximation of two-point functions. While here, as in \cite{Balasubramanian:2012tu}, we will focus on timelike two-point functions in order to be able to derive the spectral function, spacelike Green's functions as probes of thermality have been considered in \cite{AbajoArrastia:2010yt,Balasubramanian:2011ur,Aparicio:2011zy}. Spacelike geodesics connecting two equal time boundary points separated by a distance $\ell$ are conjectured to also compute the entanglement entropy of a spatial region of size $\ell$ in the boundary two-dimensional CFT \cite{Ryu:2006bv,Hubeny:2007xt,AbajoArrastia:2010yt,Balasubramanian:2011ur,Hubeny:2013hz}.  Other non-local probes of thermalisation derived from the entanglement entropy, namely the mutual and tripartite information, were discussed in Federico Galli's talk at this workshop \cite{Bernamonti:2012xv} and are studied in \cite{Balasubramanian:2011at,Allais:2011ys}.

%%%%%%%%%%%%%%%%%%%%%%%%%%%%%%%%%%%%%%%%%%%%%%%%%%%%%%%%%%%%%%%%%%%%%%%%%

\section{Geodesic approximation of two-point functions}\label{Geodesic approximation of two-point functions}

Consider a scalar operator $\mathcal{O}(t,x)$ with conformal dimension $\Delta$ in the dual CFT. The time-ordered two-point function $\langle T\{\mathcal{O}(t_{2},x_{2})\mathcal{O}(t_{1},x_{1})\}\rangle$ is given by a path integral over all paths $\mathcal{P}$ that connect the two insertion points $(t_{1},x_{1})$ and $(t_{2},x_{2})$ on the boundary:
\begin{equation}
\langle T\{\mathcal{O}(t_{2},x_{2})\mathcal{O}(t_{1},x_{1})\}\rangle=\int\mathcal{D}\mathcal{P}\,e^{-\Delta\mathcal{L}(\mathcal{P})}
\qquad\mbox{with}\qquad
\mathcal{L}(\mathcal{P})\equiv\int_{\mathcal{P}}\sqrt{g_{\mu\nu}\frac{dx^{\mu}}{d\lambda}\frac{dx^{\nu}}{d\lambda}}\mbox{d}\lambda\,.
\end{equation}
For large $\Delta$, we can use a saddle point approximation, in which the sum over all paths can be approximated by a sum over all geodesics connecting the boundary endpoints \cite{Balasubramanian:1999zv}
\begin{equation}
\langle T\{\mathcal{O}(t_{2},x_{2})\mathcal{O}(t_{1},x_{1})\}\rangle\sim\sum_{\rm{geodesics}}e^{-\Delta\mathcal{L}}\,,
\end{equation}
where ${\cal L}$ denotes the geodesic length. However, due to contributions near the asymptotically AdS boundary, the geodesic length between two boundary points contains a universal divergence and needs to be renormalized. Throughout this section, we will define a renormalized length as $\delta\mathcal{L}\equiv\mathcal{L}-2\ln r_{0}$, in terms of the bulk cut-off $r_{0}$, by removing the divergent part of the geodesic length in pure AdS. The renormalized two point function can then be approximated by
\begin{equation}
\langle T\{\mathcal{O}(t_{2},x_{2})\mathcal{O}(t_{1},x_{1})\}\rangle_{\rm ren}\sim e^{-\Delta \delta\mathcal{L} }\,,
\end{equation}
where $\delta\mathcal{L}$ is the renormalised length of the geodesic that connects the points $(t_{1},x_{1})$ and $(t_{2},x_{2})$ on the boundary.

%%%%%%%%%%%%%%%%%%%%%%%%%%%%%%%%%%%%%%%%%%%%%%%%%%%%%%%%%%%%%%%%%%%%%%%%%

\subsection{Continuation to Euclidean signature}

While for static spacetimes a Wick rotation to Euclidean signature is straightforward, this is not the case for time-dependent geometries as the AdS-Vaidya background (\ref{eq:Vaidyametric}). Therefore we perform a non-standard analytic continuation of the metric first proposed in  \cite{Balasubramanian:2012tu}. Let $E>R_{2}>R_{1}>0$ and consider the auxiliary `spacelike' Vaidya metric
\begin{eqnarray}\label{eq:spacelikeVaidya}
ds^{2}=-[(r^{2}-R_{1}^{2})-\theta(v/E)(R_{2}^{2}-R_{1}^{2})]&&dv^{2}+\frac{2Edvdr}{\sqrt{(r^{2}-R_{1}^{2})-\theta(v/E)(R_{2}^{2}-R_{1}^{2})+E^{2}}} \nonumber \\
&&+\frac{dr^{2}}{(r^{2}-R_{1}^{2})-\theta(v/E)(R_{2}^{2}-R_{1}^{2})+E^{2}}+r^{2}dx\,.
\end{eqnarray}
Through the coordinate transformation
\begin{equation}
v=\left\{\begin{array}{ll}
t-\frac{1}{R_{2}}\mbox{arctanh}\left(\frac{R_{2}}{E}\sqrt{1+\frac{(E^{2}-R_{2}^{2})}{r^{2}}}\right)+\frac{1}{R_{2}}\mbox{arctanh}\left(\frac{R_{2}}{E}\right) & \mbox{ for } v>0 \\
t-\frac{1}{R_{1}}\mbox{arctanh}\left(\frac{R_{1}}{E}\sqrt{1+\frac{(E^{2}-R_{1}^{2})}{r^{2}}}\right)+\frac{1}{R_{1}}\mbox{arctanh}\left(\frac{R_{1}}{E}\right) & \mbox{ for } v<0
\end{array}\right.\,,
\end{equation}
we recover the BTZ metric (\ref{eq:BTZ2metric}) with radius $R_{2}$ for $v>0$ and BTZ metric (\ref{eq:BTZ1metric}) with radius $R_{1}$ for $v<0$. The coordinate transformations are such that $v=0$ describes the geodesic of an infalling shell of `spacelike' particles in the BTZ spacetime. Observe that for $E>0$, $\theta(v/E)=\theta(v)$, so that in the $E\rightarrow\infty$ limit (\ref{eq:spacelikeVaidya}) reduces to the Lorentzian `null' Vaidya metric (\ref{eq:Vaidyametric}).

On the Lorentzian `spacelike' Vaidya metric (\ref{eq:spacelikeVaidya}), we can now perform an analytic continuation on the time coordinate $w=iv$, as well as on the parameter $Q=iE$. Without loss of generality, we can take $Q>0$ and find the metric
\begin{eqnarray}
ds^{2}=[(r^{2}-R_{1}^{2})-\theta(w/Q)(R_{2}^{2}-R_{1}^{2})]&&dw^{2}-\frac{2Qdwdr}{\sqrt{(r^{2}-R_{1}^{2})-\theta(w/Q)(R_{2}^{2}-R_{1}^{2})-Q^{2}}} \nonumber \\
&&+\frac{dr^{2}}{(r^{2}-R_{1}^{2})-\theta(w/Q)(R_{2}^{2}-R_{1}^{2})-Q^{2}}+r^{2}dx^{2}\,,
\end{eqnarray}
where as before $\theta(w/Q)=\theta(w)$. Note that the radial coordinate $r$ now runs from $\sqrt{Q^{2}+R_{2}^{2}}$ to $\infty$. Letting
\begin{equation}\label{eq:finout}
f(r)^{2}\equiv\left\{\begin{array}{ll}
f_{in}(r)^{2}=r^{2}-Q^{2}-R_{1}^{2} & \mbox{ for } w<0 \\
f_{out}(r)^{2}=r^{2}-Q^{2}-R_{2}^{2} & \mbox{ for } w>0
\end{array}\right.\,,
\end{equation}
the metric becomes
\begin{equation}
ds^{2}=f(r)^{2}dw^{2}+\left(Qdw-\frac{dr}{f(r)}\right)^{2}+r^{2}dx^{2}\,,
\end{equation}
which is manifestly positive.

%%%%%%%%%%%%%%%%%%%%%%%%%%%%%%%%%%%%%%%%%%%%%%%%%%%%%%%%%%%%%%%%%%%%%%%%%

\subsection{Geodesic length in Euclidean Vaidya}

We now compute the length of the geodesics that connect the points $(r_{0},\tau_{1},x_{1})$ and $(r_{0},\tau_{2},x_{2})$, where $r_{0}$ denotes the location of the regularised AdS boundary. We will assume $\Delta\tau=\tau_{2}-\tau_{1}>0$ and, for computational simplicity, mostly focus on equal space geodesics with $\Delta x=x_{2}-x_{1}=0$. The geodesics in Euclidean Vaidya consist of piecewise geodesic curves in BTZ. Above the shell ($w>0$), we have the geodesics \cite{Balasubramanian:2012tu}
\begin{eqnarray}
x_{\pm}(r)&=&x_{0}\pm\frac{1}{R_{2}}\mbox{arctanh}\left(\frac{\Gamma_{-}}{\Gamma_{+}}\sqrt{\frac{r^{2}-\Gamma_{+}^{2}}{r^{2}-\Gamma_{-}^{2}}}\right)\,, \label{eq:xpm}\\
w_{\pm}(r)&=&\tau_{0}\pm\frac{1}{R_{2}}\mbox{arctan}\left(\sqrt{\frac{R_{2}^{2}-\Gamma_{-}^{2}}{\Gamma_{+}^{2}-R_{2}^{2}}}\sqrt{\frac{r^{2}-\Gamma_{+}^{2}}{r^{2}-\Gamma_{-}^{2}}}\right) \nonumber \\
&&\qquad\qquad\qquad+\frac{1}{R_{2}}\mbox{arctan}\left(\frac{R_{2}}{Q}\sqrt{1-\frac{(Q^{2}+R_{2}^{2})}{r^{2}}}\right)-\frac{1}{R_{2}}\mbox{arctan}\left(\frac{R_{2}}{Q}\right)\,, \\
\lambda_{\pm}(r)&=&\lambda_{0}\pm\mbox{arccosh}\left(\sqrt{\frac{r^{2}-\Gamma_{-}^{2}}{\Gamma_{+}^{2}-\Gamma_{-}^{2}}}\right)\,,
\end{eqnarray}
and below the shell ($w<0$), we have the geodesics
\begin{eqnarray}
\bar{x}_{\pm}(r)&=&\bar{x}_{0}\pm\frac{1}{R_{1}}\mbox{arctanh}\left(\frac{\bar{\Gamma}_{-}}{\bar{\Gamma}_{+}}\sqrt{\frac{r^{2}-\bar{\Gamma}_{+}^{2}}{r^{2}-\bar{\Gamma}_{-}^{2}}}\right)\,, \\
\bar{w}_{\pm}(r)&=&\bar{\tau}_{0}\pm\frac{1}{R_{1}}\mbox{arctan}\left(\sqrt{\frac{R_{1}^{2}-\bar{\Gamma}_{-}^{2}}{\bar{\Gamma}_{+}^{2}-R_{1}^{2}}}\sqrt{\frac{r^{2}-\bar{\Gamma}_{+}^{2}}{r^{2}-\bar{\Gamma}_{-}^{2}}}\right) \nonumber \\
&&\qquad\qquad\qquad+\frac{1}{R_{1}}\mbox{arctan}\left(\frac{R_{1}}{Q}\sqrt{1-\frac{(Q^{2}+R_{1}^{2})}{r^{2}}}\right)-\frac{1}{R_{1}}\mbox{arctan}\left(\frac{R_{1}}{Q}\right)\,, \\
\bar{\lambda}_{\pm}(r)&=&\bar{\lambda}_{0}\pm\mbox{arccosh}\left(\sqrt{\frac{r^{2}-\bar{\Gamma}_{-}^{2}}{\bar{\Gamma}_{+}^{2}-\bar{\Gamma}_{-}^{2}}}\right)\,. \label{eq:lambdapm}
\end{eqnarray}
Here $\pm$ denote two separate branches (appearing above and below the turning points of BTZ geodesics) and the parameters appearing in these expressions satisfy the conditions $0\leqslant\Gamma_{-}^{2}\leqslant R_{2}^{2}<\Gamma_{+}^{2}$ and $0\leqslant\bar{\Gamma}_{-}^{2}\leqslant R_{1}^{2}<\bar{\Gamma}_{+}^{2}$. We can now distinguish three cases.
\begin{itemize}
\item[1)] If both endpoints are above the shell ($\tau_{2}>\tau_{1}>0$), then there is a geodesic connecting them which is entirely above the shell. Any additional geodesics would require the existence of geodesics below the shell in BTZ that connect two points on the shell, and, as we show below, these do not exist.
\item[2)] If both endpoints are below the shell ($0>\tau_{2}>\tau_{1}$), then there is a geodesic connecting them which is entirely below the shell. Any additional geodesics would require the existence of geodesics above the shell in BTZ that connect two points on the shell, and again these do not exist.
\end{itemize}
The geodesic length is then given by the BTZ result, derived for example in the Appendix of \cite{Balasubramanian:2012tu},
\begin{equation}\label{eq:BTZlength}
\delta\mathcal{L}=\ln\left[\frac{4}{R^{2}}\left(\sin^{2}\left(\frac{R\Delta\tau}{2}\right)+\sinh^{2}\left(\frac{R\Delta x}{2}\right)\right)\right]\,,
\end{equation}
where $R$ is respectively $R_{2}$ and $R_{1}$ in case 1) and 2). 

The existence of geodesics that have two endpoints on the shell, and that are above (or below) the shell would require the existence of a local maximum (or minimum) of $w(r)$ (or $\bar{w}(r)$). We note that $\frac{dw_{+}}{dr}(r_{\odot})=0$ has no real solutions, so no geodesic in BTZ can have a local maximum. On the other hand $\frac{d\bar{w}_{-}}{dr}(r_{\odot})=0$ can be solved by $r_{\odot}=\frac{Q\Gamma_{+}\Gamma_{-}}{\sqrt{Q^{2}R_{1}^{2}-(\bar{\Gamma}_{+}^{2}-R_{1}^{2})(R_{1}^{2}-\bar{\Gamma}_{-}^{2})}}$. Equal space geodesics as we consider here, all have $\bar{\Gamma}_{-}=0$ and thus no local minimum.
\begin{itemize}
\item[3)] Finally we consider geodesics with one endpoint above and one below the shell ($\tau_{2}>0>\tau_{1}$). They cross the shock wave once and hence we need to determine how they refract at the shell location in order to compute their length.
\end{itemize}
The conditions that need to be imposed for the geodesics at the shell ($w=0$) were derived in \cite{Balasubramanian:2012tu}. They are given by the following refraction law
\begin{eqnarray}\label{eq:refractionlaw}
\left.\frac{1}{f^2\frac{dx}{dr}}\left(Qf\frac{dw}{dr}-1\right)\right|_{in}&=&\left.\frac{1}{f^2\frac{dx}{dr}}\left(Qf\frac{dw}{dr}-1\right)\right|_{out}\,, \nonumber \\
\left.\frac{1}{\left(f\frac{dx}{dr}\right)^{2}}\left[f^4\left(\frac{dw}{dr}\right)^{2}+\left(Qf\frac{dw}{dr}-1\right)^{2}\right]\right|_{in}&=&\left.\frac{1}{\left(f\frac{dx}{dr}\right)^{2}}\left[f^4\left(\frac{dw}{dr}\right)^{2}+\left(Qf\frac{dw}{dr}-1\right)^{2}\right]\right|_{out}\,,
\qquad\quad
\end{eqnarray}
which is supplemented by the continuity conditions $x_{in}(r_{*})=x_{out}(r_{*})$ and $w_{in}(r_{*})=w_{out}(r_{*})=0$, where $r_{*}$ is the value of the radial coordinate at which the geodesic reaches the shell. 

\begin{figure}[!h]
\centering
\includegraphics[width=8cm]{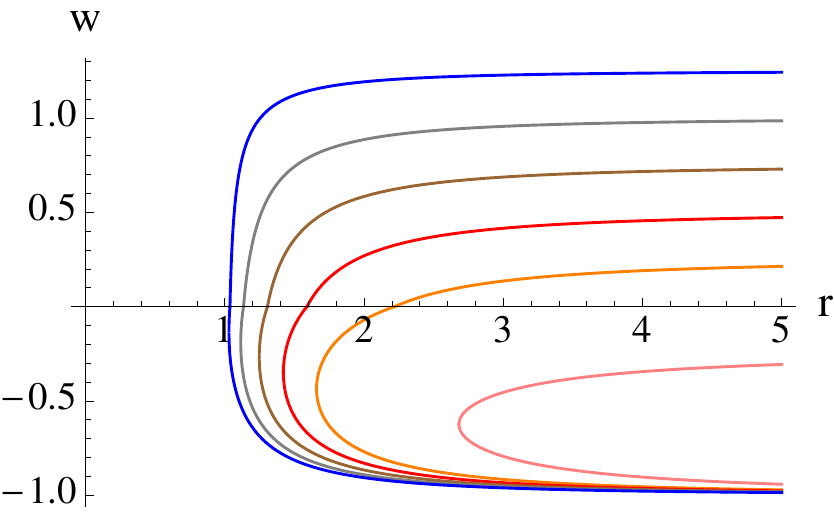}
\caption{\emph{ Equal space geodesics in the Euclidean shell background for $R_1 =0.5$, $R_2 =1$, $Q=0.25$, $\tau_1 = -1$ and, increasing from the bottom up, $\tau_2 = -0.25,0.25, 0.5, 0.75,1,1.25$. }}\label{fig:geodesics}
\end{figure}

The boundary conditions at the AdS boundary are
\begin{equation}
\left\{\begin{array}{ll}
x_{+}(r\rightarrow\infty)=x_{2} \\
w_{+}(r\rightarrow\infty)=\tau_{2} \\
\bar{x}_{-}(r\rightarrow\infty)=x_{1} \\
\bar{w}_{-}(r\rightarrow\infty)=\tau_{1}
\end{array}\right.
\qquad\Rightarrow\qquad
\left\{\begin{array}{ll}
x_{2}=x_{0}+\frac{1}{R_{2}}\mbox{arctanh}\left(\frac{\Gamma_{-}}{\Gamma_{+}}\right) \\
\tau_{2}=\tau_{0}+\frac{1}{R_{2}}\mbox{arctan}\left(\sqrt{\frac{(R_{2}^{2}-\Gamma_{-}^{2})}{(\Gamma_{+}^{2}-R_{2}^{2})}}\right) \\
x_{1}=\bar{x}_{0}-\frac{1}{R_{1}}\mbox{arctanh}\left(\frac{\bar{\Gamma}_{-}}{\bar{\Gamma}_{+}}\right) \\
\tau_{1}=\bar{\tau}_{0}-\frac{1}{R_{1}}\mbox{arctan}\left(\sqrt{\frac{(R_{1}^{2}-\bar{\Gamma}_{-}^{2})}{(\bar{\Gamma}_{+}^{2}-\bar{R}^{2})}}\right)
\end{array}\right.
\end{equation}
and the continuity conditions 
\begin{equation}
x_{0}+\frac{1}{R_{2}}\mbox{arctanh}\left(\frac{\Gamma_{-}}{\Gamma_{+}}\sqrt{\frac{r_{*}^{2}-\Gamma_{+}^{2}}{r_{*}^{2}-\Gamma_{-}^{2}}}\right)=\bar{x}_{0}+\frac{1}{R_{1}}\mbox{arctanh}\left(\frac{\bar{\Gamma}_{-}}{\bar{\Gamma}_{+}}\sqrt{\frac{r_{*}^{2}-\bar{\Gamma}_{+}^{2}}{r_{*}^{2}-\bar{\Gamma}_{-}^{2}}}\right)\,,
\end{equation}
\begin{eqnarray}
0&=&\tau_{0}+\frac{1}{R_{2}}\mbox{arctan}\left(\sqrt{\frac{R_{2}^{2}-\Gamma_{-}^{2}}{\Gamma_{+}^{2}-R_{2}^{2}}}\sqrt{\frac{r_{*}^{2}-\Gamma_{+}^{2}}{r_{*}^{2}-\Gamma_{-}^{2}}}\right) \nonumber \\
&&\qquad\qquad\qquad+\frac{1}{R_{2}}\mbox{arctan}\left(\frac{R_{2}}{Q}\sqrt{1-\frac{(Q^{2}+R_{2}^{2})}{r_{*}^{2}}}\right)-\frac{1}{R_{2}}\mbox{arctan}\left(\frac{R_{2}}{Q}\right)\,, \\
0&=&\bar{\tau}_{0}+\frac{1}{R_{1}}\mbox{arctan}\left(\sqrt{\frac{R_{1}^{2}-\bar{\Gamma}_{-}^{2}}{\bar{\Gamma}_{+}^{2}-R_{1}^{2}}}\sqrt{\frac{r_{*}^{2}-\bar{\Gamma}_{+}^{2}}{r_{*}^{2}-\bar{\Gamma}_{-}^{2}}}\right) \nonumber \\
&&\qquad\qquad\qquad+\frac{1}{R_{1}}\mbox{arctan}\left(\frac{R_{1}}{Q}\sqrt{1-\frac{(Q^{2}+R_{1}^{2})}{r_{*}^{2}}}\right)-\frac{1}{R_{1}}\mbox{arctan}\left(\frac{R_{1}}{Q}\right)\,.
\end{eqnarray}
In terms of the solutions \eqref{eq:xpm}-\eqref{eq:lambdapm}, the refraction conditions read
\begin{equation}
\frac{\Gamma_{-}^{2}\Gamma_{+}^{2}}{R_{2}^{2}}=\frac{\bar{\Gamma}_{-}^{2}\bar{\Gamma}_{+}^{2}}{R_{1}^{2}}\,,
\end{equation}
\begin{eqnarray} \label{eq:rstar}
&&\frac{1}{(r_{*}^{2}-R_{2}^{2})R_{2}}\left(R_{2}\sqrt{(r_{*}^{2}-\Gamma_{-}^{2})(r_{*}^{2}-\Gamma_{+}^{2})}-Qr_{*}\frac{\sqrt{(R_{2}^{2}-\Gamma_{-}^{2})(\Gamma_{+}^{2}-R_{2}^{2})}}{\sqrt{r_{*}^{2}-Q^{2}-R_{2}^{2}}}\right) \nonumber \\
&&\qquad=\frac{1}{(r_{*}^{2}-R_{1}^{2})R_{1}}\left(R_{1}\sqrt{(r_{*}^{2}-\bar{\Gamma}_{-}^{2})(r_{*}^{2}-\bar{\Gamma}_{+}^{2})}-Qr_{*}\frac{\sqrt{(R_{1}^{2}-\bar{\Gamma}_{-}^{2})(\bar{\Gamma}_{+}^{2}-R_{1}^{2})}}{\sqrt{r_{*}^{2}-Q^{2}-R_{1}^{2}}}\right)\,.
\end{eqnarray}
Altogether, these are nine conditions, for nine unknowns ($x_{0}$, $\tau_{0}$, $\bar{x}_{0}$, $\bar{\tau}_{0}$, $\Gamma_{+}$, $\Gamma_{+}$, $\bar{\Gamma}_{+}$, $\bar{\Gamma}_{-}$ and $r_{*}$). The renormalised length of the geodesic is given by
\begin{equation}
\delta\mathcal{L}=\lim_{r_{0}\rightarrow\infty}\left(\lambda_{+}(r_{0})-\bar{\lambda}_{-}(r_{0})-2\ln(r_{0})\right)=\lambda_{0}-\bar{\lambda}_{0}+\frac{1}{2}\ln\left(\frac{4}{\Gamma_{+}^{2}-\Gamma_{-}^{2}}\right)+\frac{1}{2}\ln\left(\frac{4}{\bar{\Gamma}_{+}^{2}-\bar{\Gamma}_{-}^{2}}\right)\,,
\end{equation}
which, using the continuity condition $\lambda_{+}(r_{*})=\bar{\lambda}_{+}(r_{*})$, can be written as
\begin{equation}
\delta\mathcal{L}=\ln\left(\frac{4}{(\bar{\Gamma}_{+}^{2}-\bar{\Gamma}_{-}^{2})}\frac{\sqrt{r_{*}^{2}-\bar{\Gamma}_{-}^{2}}+\sqrt{r_{*}^{2}-\bar{\Gamma}_{+}^{2}}}{\sqrt{r_{*}^{2}-\Gamma_{-}^{2}}+\sqrt{r_{*}^{2}-\Gamma_{+}^{2}}}\right)\,.
\end{equation}
We will focus on equal space correlators, {\it i.e.} $\Delta x=x_{2}-x_{1}=0$. Because of the refraction condition, $\frac{dx}{dr}$ does not change sign at the shell. Also the geodesics in BTZ have no local extremum of $x(r)$. Therefore, we must set $\Gamma_{-}=\bar{\Gamma}_{-}=0$. It then follows that $x_{0}=\bar{x}_{0}=x_{1}=x_{2}$. We are thus left with five conditions for five unknowns. After solving this system, we find that the renormalised geodesic length is given by
\begin{equation}
\delta\mathcal{L}=\ln\left(\frac{4}{(r_{*}-\cos(\bar{\gamma})\sqrt{r_{*}^{2}-R_{1}^{2}})(r_{*}+\cos(\gamma)\sqrt{r_{*}^{2}-R_{2}^{2}})}\right)\,,
\end{equation}
where the unknown $r_{*}$ is implicitly determined by the relation\footnote{The expressions (\ref{eq:RstarSolution1})-(\ref{eq:RstarSolution3}) were obtained assuming $0\leqslant\gamma(r_{*})\leqslant\pi/2$ and $0\leqslant\bar{\gamma}(r_{*})\leqslant\pi/2$. Outside this range, similar expressions can be derived from \eqref{eq:rstar}.}
\begin{equation}\label{eq:RstarSolution1}
\frac{1}{\sqrt{r_{*}^{2}-R_{2}^{2}}}\left(\cos(\gamma)-\frac{Q\sin(\gamma)}{\sqrt{r_{*}^{2}-Q^{2}-R_{2}^{2}}}\right)=\frac{1}{\sqrt{r_{*}^{2}-R_{1}^{2}}}\left(\cos(\bar{\gamma})-\frac{Q\sin(\bar{\gamma})}{\sqrt{r_{*}^{2}-Q^{2}-R_{1}^{2}}}\right)\,,
\end{equation}
and we have used the expressions
\begin{equation}\label{eq:RstarSolution2}
\sin(\gamma(r_{*}))=\frac{R_{2}\sin(R_{2}\Omega(r_{*}))}{r_{*}-\cos(R_{2}\Omega(r_{*}))\sqrt{r_{*}^{2}-R_{2}^{2}}}\,,
\end{equation}
and
\begin{equation}\label{eq:RstarSolution3}
\sin(\bar{\gamma}(r_{*}))=-\frac{R_{1}\sin(R_{1}\bar{\Omega}(r_{*}))}{r_{*}-\cos(R_{1}\bar{\Omega}(r_{*}))\sqrt{r_{*}^{2}-R_{1}^{2}}}\,,
\end{equation}
with
\begin{equation}
\Omega(r_{*})=\tau_{2}+\frac{1}{R_{2}}\mbox{arctan}\left(\frac{R_{2}}{Q}\sqrt{1-\frac{(Q^{2}+R_{2}^{2})}{r_{*}^{2}}}\right)-\frac{1}{R_{2}}\mbox{arctan}\left(\frac{R_{2}}{Q}\right)\,,
\end{equation}
and
\begin{equation}
\bar{\Omega}(r_{*})=\tau_{1}+\frac{1}{R_{1}}\mbox{arctan}\left(\frac{R_{1}}{Q}\sqrt{1-\frac{(Q^{2}+R_{1}^{2})}{r_{*}^{2}}}\right)-\frac{1}{R_{1}}\mbox{arctan}\left(\frac{R_{1}}{Q}\right)\,.
\end{equation}
The parameters $\Gamma_+$ and $\bar \Gamma_+$ in the geodesics \eqref{eq:xpm}-\eqref{eq:lambdapm} are given by the formulas $\Gamma_{+}^{2}(r_{*})=R_{2}^{2}+(r_{*}^{2}-R_{2}^{2})\sin^{2}(\gamma(r_{*}))$ and $\bar{\Gamma}_{+}^{2}(r_{*})=R_{1}^{2}+(r_{*}^{2}-R_{1}^{2})\sin^{2}(\bar{\gamma}(r_{*}))$. In Fig.~\ref{fig:geodesics}, we plot the profile of a sample of equal space geodesics in the Euclidean AdS-Vaidya geometry. 

If we perform the double Wick rotation $\tau_{1}=it_{1}$, $\tau_{2}=it_{2}$ and $Q=iE$ and take the limit $E\rightarrow\infty$, in order to recover the `lightlike' Vaidya result, we finally find
\begin{equation}\label{GeodesicLengthVaidya}
\delta\mathcal{L}=\ln\left[\left(\frac{2}{R_{1}}\cosh\left(\frac{R_{2}t_{2}}{2}\right)\sinh\left(\frac{R_{1}t_{1}}{2}\right)-\frac{2}{R_{2}}\sinh\left(\frac{R_{2}t_{2}}{2}\right)\cosh\left(\frac{R_{1}t_{1}}{2}\right)\right)^{2}\right]\,,
\end{equation}
which reduces to the expression obtained in \cite{Balasubramanian:2012tu} in the $R_1 =0$ case. 

%%%%%%%%%%%%%%%%%%%%%%%%%%%%%%%%%%%%%%%%%%%%%%%%%%%%%%%%%%%%%%%%%%%%%%%%%

\subsection{Two-point functions}

%%%%%%%%%%%%%%%%%%%%%%%%%%%%%%%%%%%%%%%%%%%%%%%%%%%%%%%%%%%%%%%%%%%%%%%%%

\subsubsection{Equilibrium CFT}

The Euclidean two-point function associated to a CFT in thermal equilibrium at a temperature $\theta=R/(2\pi)$ can be computed in the geodesic approximation from the renormalised geodesic length in Euclidean BTZ (\ref{eq:BTZlength}). Remember that, in the geodesic approximation, the renormalised Euclidean two-point function in the dual CFT equals
\begin{equation}
\langle\mathcal{O}(\tau_{2},x_{2})\mathcal{O}(\tau_{1},x_{1})\rangle_{ren}\sim e^{-\Delta \delta\mathcal{L} }\,.
\end{equation}
From (\ref{eq:BTZlength}), we then find the (Euclidean) thermal two-point function
\begin{equation}
G_{E}^{\mbox{thermal}}(\tau_{2},x_{2};\tau_{1},x_{1})=\frac{1}{\left[\frac{4}{R^{2}}\left(\sin^{2}\left(\frac{R\Delta\tau}{2}\right)+\sinh^{2}\left(\frac{R\Delta x}{2}\right)\right)\right]^{\Delta}}\,.
\end{equation}
Despite the fact that we have used a geodesic approximation, the result for the thermal two-point function is exact, being fully constrained by conformal invariance (apart from an overall scaling). Note that we can obtain the vacuum result by taking $R\rightarrow0$
\begin{equation}
G_{E}^{\mbox{vacuum}}(\tau_{2},x_{2};\tau_{1},x_{1})=\lim_{R\rightarrow0}G_{E}^{\mbox{thermal}}(\tau_{2},x_{2};\tau_{1},x_{1})=\frac{1}{\left(\Delta\tau^{2}+\Delta x^{2}\right)^{\Delta}}\,.
\end{equation}
It is well known that by Wick rotating the Euclidean two-point function, one finds the time-ordered (Feynman) two-point function. Therefore let $\tau=\lim_{\varepsilon\rightarrow0}e^{i(\frac{\pi}{2}-\varepsilon)}t\simeq it$, such that $\tau^{2}=-t^{2}+i\epsilon$ and \begin{equation}
iG_{F}(t_{2},x_{2};t_{1},x_{1})=G_{E}(it_{2},x_{2};it_{1},x_{1})\,.
\end{equation}
The retarded two-point function can be obtained from the time-ordered one by applying the relation (\ref{eq:RelationFeynmanRetarded}). For non-integer scaling dimension $\Delta$,\footnote{The case of integer $\Delta$ is discussed in detail in \cite{Balasubramanian:2012tu}.} this leads to
\begin{equation}
iG_{F}^{\mbox{thermal}}(t_{2},x_{2};t_{1},x_{1})=\frac{\theta\left[(\Delta x)^{2}-(\Delta t)^{2}\right]}{\left[\frac{4}{R^{2}}\left(\sinh^{2}\left(\frac{R\Delta x}{2}\right)-\sinh^{2}\left(\frac{R\Delta t}{2}\right)\right)\right]^{\Delta}}+\frac{\theta\left[(\Delta t)^{2}-(\Delta x)^{2}\right]e^{-i\pi\Delta}}{\left[\frac{4}{R^{2}}\left(\sinh^{2}\left(\frac{R\Delta t}{2}\right)-\sinh^{2}\left(\frac{R\Delta x}{2}\right)\right)\right]^{\Delta}}\,, \label{eq:GFthermal}
\end{equation}
and
\begin{equation}
G_{R}^{\mbox{thermal}}(t_{2},x_{2};t_{1},x_{1})=-2\sin\left(\pi\Delta\right)\theta\left(\Delta t\right)\frac{\theta\left[(\Delta t)^{2}-(\Delta x)^{2}\right]}{\left[\frac{4}{R^{2}}\left(\sinh^{2}\left(\frac{R\Delta t}{2}\right)-\sinh^{2}\left(\frac{R\Delta x}{2}\right)\right)\right]^{\Delta}}\,. \label{eq:GRthermal}
\end{equation}

%%%%%%%%%%%%%%%%%%%%%%%%%%%%%%%%%%%%%%%%%%%%%%%%%%%%%%%%%%%%%%%%%%%%%%%%%

\subsubsection{Thermalising CFT}

In an analogous way, from (\ref{GeodesicLengthVaidya}) we can then find (the geodesic approximation of) the Feynman and retarded two-point functions in the thermalizing CFT dual to three-dimensional Vaidya:
\begin{eqnarray}
iG_{F}(t_{2},x;t_{1},x)&=&\left(\frac{e^{-i\pi\Delta}\theta(-t_{1})\theta(-t_{2})}{\left|\frac{2}{R_{1}}\sinh\left(\frac{R_{1}}{2}(t_{2}-t_{1})\right)\right|^{2\Delta}}\right)+\left(\frac{e^{-i\pi\Delta}\theta(t_{1})\theta(t_{2})}{\left|\frac{2}{R_{2}}\sinh\left(\frac{R_{2}}{2}(t_{2}-t_{1})\right)\right|^{2\Delta}}\right) \nonumber \\
&&+\left(\frac{e^{-i\pi\Delta}\theta(t_{1})\theta(-t_{2})}{\left|\frac{2}{R_{1}}\cosh\left(\frac{R_{2}t_{2}}{2}\right)\sinh\left(\frac{R_{1}t_{1}}{2}\right)-\frac{2}{R_{2}}\sinh\left(\frac{R_{2}t_{2}}{2}\right)\cosh\left(\frac{R_{1}t_{1}}{2}\right)\right|^{2\Delta}}\right) \nonumber \\
&&+\left(\frac{e^{-i\pi\Delta}\theta(-t_{1})\theta(t_{2})}{\left|\frac{2}{R_{1}}\cosh\left(\frac{R_{2}t_{1}}{2}\right)\sinh\left(\frac{R_{1}t_{2}}{2}\right)-\frac{2}{R_{2}}\sinh\left(\frac{R_{2}t_{1}}{2}\right)\cosh\left(\frac{R_{1}t_{2}}{2}\right)\right|^{2\Delta}}\right)\,, \label{eq:GFVaydia}
\end{eqnarray}
and
\begin{eqnarray}
G_{R}(t_{2},x;t_{1},x)&=&-2\sin(\pi\Delta)\theta(t_{2}-t_{1})\left\{\left(\frac{\theta(-t_{1})\theta(-t_{2})}{\left|\frac{2}{R_{1}}\sinh\left(\frac{R_{1}}{2}(t_{2}-t_{1})\right)\right|^{2\Delta}}\right)+\left(\frac{\theta(t_{1})\theta(t_{2})}{\left|\frac{2}{R_{2}}\sinh\left(\frac{R_{2}}{2}(t_{2}-t_{1})\right)\right|^{2\Delta}}\right)\right. \nonumber \\
&&\qquad\qquad\left.+\left(\frac{\theta(-t_{1})\theta(t_{2})}{\left|\frac{2}{R_{1}}\cosh\left(\frac{R_{2}t_{1}}{2}\right)\sinh\left(\frac{R_{1}t_{2}}{2}\right)-\frac{2}{R_{2}}\sinh\left(\frac{R_{2}t_{1}}{2}\right)\cosh\left(\frac{R_{1}t_{2}}{2}\right)\right|^{2\Delta}}\right)\right\}\,. \label{eq:GRVaydia}
\end{eqnarray}
In the limit $R_{2}\rightarrow R_{1}$ of equal initial and final temperatures, these expressions reduce to the equilibrium ones, (\ref{eq:GFthermal}) and (\ref{eq:GRthermal}). Also, in the limit of zero initial temperature, $R_{1}\rightarrow 0$, we recover the expressions for a thermalising CFT that were presented in \cite{Balasubramanian:2012tu}.

%%%%%%%%%%%%%%%%%%%%%%%%%%%%%%%%%%%%%%%%%%%%%%%%%%%%%%%%%%%%%%%%%%%%%%%%%

\section{Spectral function and effective temperature in thermalising CFT}\label{Spectral function and effective temperature in thermalising CFT}

\begin{figure}[h!]
\centering
\includegraphics[width=7.6cm]{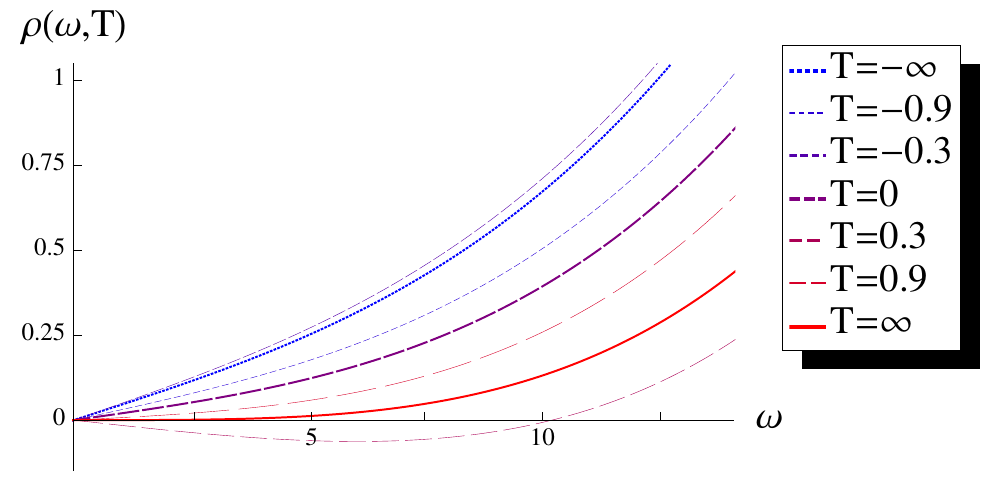}\includegraphics[width=7.6cm]{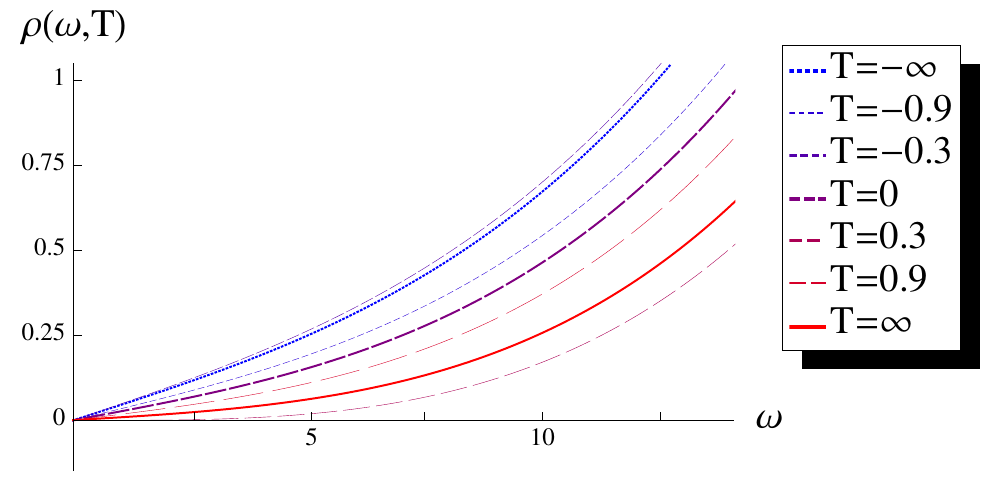}\\{\bf (A)} \hfil\hfil {\bf (B)}
\caption{\emph{The spectral function as a function of the frequency $\omega$ for different values of the average time $T$. In both plots, we have taken $\Delta=2.25$ and $R_{2}=1$. The plot {\bf (A)} has $R_{1}=0$ and plot {\bf (B)} has $R_{1}=0.5$.}}\label{fig:spectralfct}
\end{figure}

\begin{figure}[h!]
\centering
\includegraphics[width=8.7cm]{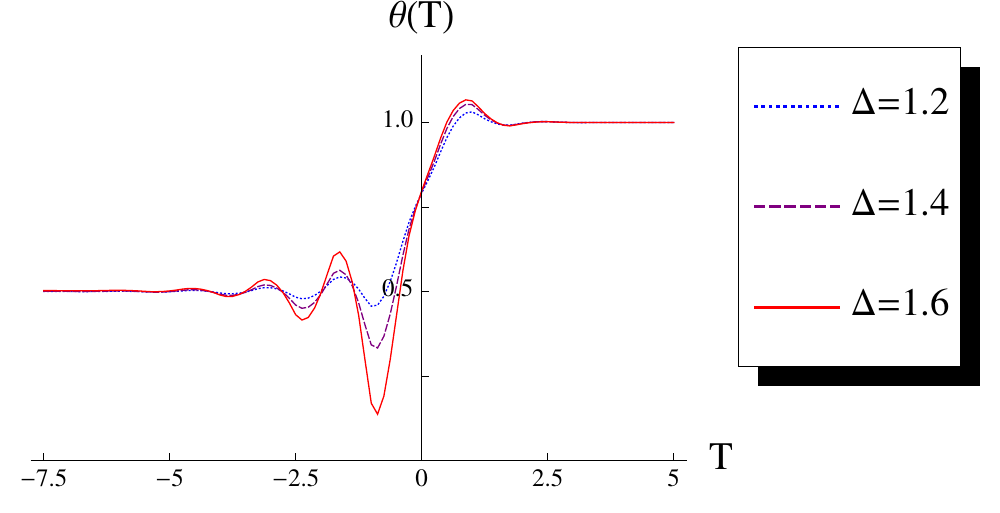}\includegraphics[width=6.3cm]{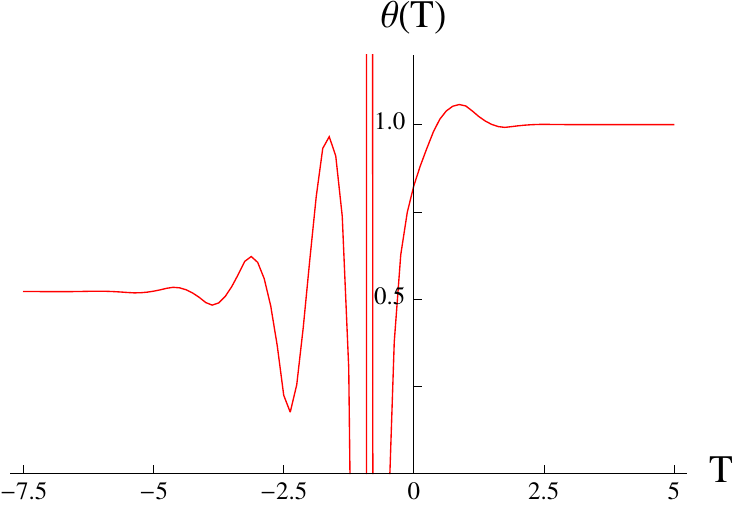}\\{\bf (A)} \hfil\hfil {\bf (B)}
\caption{\emph{The temperature as a function of average time for different values of the conformal dimension $\Delta$. In both plots, we have taken $R_{2}=1$ and $R_{1}=0.5$. In plot {\bf (A)} we have taken different values for $\Delta$ as indicated in the legend. Plot {\bf (B)} has $\Delta=2.25$.}}\label{fig:temperature}
\end{figure}

In a two-dimensional CFT in equilibrium at a temperature $\theta=R/(2\pi)$, we know the spectral function analytically \cite{Balasubramanian:2012tu,Son:2002sd}. It is obtained from the Fourier transform of the retarded two-point function (\ref{eq:GRthermal}) using equation (\ref{eq:DefSpectralFunction}):
\begin{equation}
\rho(\omega,k)=\frac{R^{2\Delta-2}}{(\Gamma(\Delta))^{2}}\left|\Gamma\left(\frac{\Delta}{2}+\frac{i(\omega+k)}{2R}\right)\Gamma\left(\frac{\Delta}{2}+\frac{i(\omega-k)}{2R}\right)\right|^{2}\sinh\left(\frac{\pi\omega}{R}\right)\,.
\end{equation}
When averaged over momenta, it becomes
\begin{equation}\label{eq:MomAvSpectralFunctionEqui}
\rho(\omega)=\int\frac{\mbox{d}k}{2\pi}\,\rho(\omega,k)=\frac{2R^{2\Delta-1}}{\Gamma(2\Delta)}\left|\Gamma\left(\Delta+\frac{i\omega}{R}\right)\right|^{2}\sinh\left(\frac{\pi\omega}{R}\right)\,.
\end{equation}
To compute the momentum average of the time-dependent spectral function (\ref{eq:DefTimeDepSpectralFunction}) in a thermalising CFT, we need to perform a numerical Fourier transform of the equal space retarded two-point function (\ref{eq:GRVaydia}). As explained in section \ref{Time-dependent two-point functions}, one can also define a notion of time-dependent temperature (\ref{eq:DefTimeDepTemperature}) from the Fourier transforms of the thermalising Feynman and retarded equal space two-point functions. The results of this analysis are shown in Figures~\ref{fig:spectralfct} and \ref{fig:temperature}.

In the distant past ($T\rightarrow-\infty$) the spectral function is given exactly by the equilibrium expression (\ref{eq:MomAvSpectralFunctionEqui}) for a temperature $\theta_{1}=R_{1}/(2\pi)$, while in the distant future ($T\rightarrow\infty$) it coincides with the equilibrium result for a temperature $\theta_{2}=R_{2}/(2\pi)$. The notion of time-dependent spectral function that we have defined interpolates smoothly between these curves. However, for a definite frequency $\omega$, $\rho(\omega, T)$ does not increase monotonically from the past limiting value to the future one, but is an oscillating function.  

For sufficiently small values of $\Delta$, the time-dependent notion of temperature that we have defined gives a smooth transition from the initial temperature to the final temperature. However, for larger values of $\Delta$ there are values of the average time for which the spectral function will have zero slope at $\omega=0$. This results in a divergence in the temperature just before the time $T=0$ as shown in the plots. Similar singular behavior of the effective temperature was also observed in \cite{Chesler:2011ds}.

%%%%%%%%%%%%%%%%%%%%%%%%%%%%%%%%%%%%%%%%%%%%%%%%%%%%%%%%%%%%%%%%%%%%%%%%%

\section*{Acknowledgments}

JV would like to thank the organizers of the Corfu Summer Institute and the XVIIIth European Workshop on String Theory for the nice workshop and the opportunity to present this work. We would also like to thank V. Balasubramanian, V. Ker\"anen, E. Keski-Vakkuri, B. M\"uller and L. Thorlacius for enjoyable collaboration. This work is supported by the FWO-Vlaanderen, Projects No.\ G.0651.11 and G.0114.10N, by the Belgian Federal Science Policy Office through the Interuniversity Attraction Pole P7/37, by the European Science Foundation Holograv Network and by the Vrije Universiteit Brussel through the Strategic Research Program ``High-Energy Physics''. AB is a Postdoctoral Researcher FWO-Vlaanderen. JV is an Aspirant FWO-Vlaanderen.

%%%%%%%%%%%%%%%%%%%%%%%%%%%%%%%%%%%%%%%%%%%%%%%%%%%%%%%%%%%%%%%%%%%%%%%%%

%%%%%%%%%%%%%%%%%%%%%%%%%%%%%%%%%%%%%%%%%%%%%%%%%%%%%%%%%%%%%%%%%%%%%%%%%

\end{document}